\documentclass{article}

\usepackage{algorithmic}
\usepackage{algorithm}
\usepackage{graphicx}
\usepackage{xcolor}
\usepackage{xspace}
\usepackage{tikz}
\usepackage{bbold}
\usepackage{anyfontsize}
\usepackage{tikzinclude}
\usepackage{fontawesome}
\usepackage{adjustbox}
\usepackage{booktabs}
\usepackage{authblk}
\usepackage{url}
\usepackage[hidelinks=true]{hyperref}

\usepackage{amsbsy}
\usepackage{amsopn}
\usepackage{amstext}
\usepackage{amsmath}    
\usepackage{amsfonts}   
\usepackage{amssymb}    
\usepackage{amsthm} 
    
\usetikzlibrary{calc}
\usetikzlibrary{positioning}
\usetikzlibrary{shapes}
\usetikzlibrary{shapes.multipart}
\usetikzlibrary{arrows.meta}
\usetikzlibrary{decorations}
\usetikzlibrary{matrix}
\usetikzlibrary{fit}

\newcommand{\etal}{\emph{et al.}\@\xspace}

\colorlet{myyellow}{yellow!80!red}
\colorlet{mygray}{black!10!white}

\newcommand{\markers}{\mathcal{M}}
\newcommand{\telomeres}{\mathcal{T}}
\newcommand{\extremities}{\mathcal{E}}
\newcommand{\cscandidates}{\mathcal{C}}
\newcommand{\family}{\mathbf{f}}
\newcommand{\extf}{\mathbf{e}}
\newcommand{\multiplicity}{\mathbf{m}}
\newcommand{\phylo}{\Gamma}
\newcommand{\w}{\mathbf{w}}
\newcommand{\dist}{\mathbf{d}}
\newcommand\id{{\textnormal{id}}}
\newcommand{\adj}{{\textnormal{adj}}}
\newcommand{\ext}{{\textnormal{ext}}}
\newcommand{\et}{\textnormal{t}}
\newcommand{\eh}{\textnormal{h}}
\newcommand{\MR}{\mathit{MR}_\circ}
\newcommand{\DCJid}{\mathrm{DCJ-ID}}

\renewcommand{\paragraph}[1]{\smallskip\emph{#1}.~~}

\newtheorem{problem}{Problem}
\newtheorem{example}{Example}

\tikzset{
    position/.style args={#1:#2 from #3}{
        at=(#3.#1), anchor=#1+180, shift=(#1:#2)
    },
    bicolor/.style 2 args={
        dashed,dash pattern=on 0.25cm off 0.25cm,#1,
        postaction={draw,dashed,dash pattern=on 0.25cm off 0.25cm,#2,dash
        phase=0.25cm}
    },
}

\title{Small Parsimony for Natural Genomes in the DCJ-Indel Model}
\author[1]{Daniel Doerr\footnote{daniel dot doerr at hhu.de}}
\author[2]{Cedric Chauve\footnote{cedric dot chauve at sfu.ca}}

\affil[1]{\mbox{Faculty of Medicine, Heinrich Heine University, D\"usseldorf, Germany}}
\affil[2]{Department of Mathematic, Simon Fraser University, Canada}

\date{}

\begin{document}

\maketitle

\begin{abstract}
Reconstructing ancestral gene orders is an important step towards understanding genome evolution. The Small Parsimony Problem (SPP) has been extensively studied in this regard. 
The problem aims at finding the gene orders at internal nodes of a given phylogenetic tree such that the overall genome rearrangement distance along the tree branches is minimized. 
However, this problem is intractable in most genome rearrangement models, especially when gene duplication and loss are considered.
In this work, we describe an Integer Linear Program algorithm to solve the SPP for natural genomes, \emph{i.e.}, genomes that contain conserved, unique, and duplicated markers. 
The evolutionary model that we consider is the DCJ-indel model that includes the Double-Cut and Join rearrangement operation and the insertion and deletion of genome segments.
We evaluate our algorithm on simulated data and show that it is able to reconstruct very efficiently and accurately ancestral gene orders in a very comprehensive evolutionary model.
\\

\end{abstract}

\section{Introduction}
\label{sec:introduction}

The reconstruction of ancestral genomes is a well-studied problem in comparative genomics, with many applications, \emph{e.g.} in evolutionary genomics~\cite{LAMPREY,ROSIDS,PINEAPPLE,ANOPHELESa} or genome assembly~\cite{ANOPHELESb,ANOPHELESc}.
Ancestral genome reconstruction methods take as input a phylogenetic tree (a species tree) together with gene orders for extant species and aim at computing the gene orders at the internal nodes of the tree, \emph{i.e.}, ancestral species, while optimizing a suitable criterion.
Since it was introduced as an algorithmic problem in~\cite{SankoffB97,SankoffB98}, in the specific case of an unrooted phylogeny with a single ancestral species (the \textit{genome median}), this problem has been actively studied and many methods have been developed.

Approaches designed to reconstruct ancestral gene orders can be widely divided into two types: homology-based and parsimony-based. Homology-based approaches were introduced in \cite{BergeronBC04} and do not consider genome rearrangements directly.
Instead, they rely on the use of \textit{conserved genomic features}, such as gene adjacencies or common intervals, associated to specific internal nodes of the species tree and obtained by the comparison of the extant gene orders. 
In a following step, these genomic features can then be assembled into larger gene orders for the considered ancestral species, often called \emph{Contiguous Ancestral Regions} (CARs)~\cite{MaZS06,ploscb/ChauveT08}.
Conversely, parsimony-based approaches are guided by the principle of minimizing the evolutionary cost, in a given genome rearrangement model, along the branches of the considered species tree. 
This approach builds upon many tractability results on the pairwise genome rearrangement distance problem \cite{FertinLR09} and the corresponding computational problem is known as the \emph{Small Parsimony Problem} (SPP). 
However, even when restricted to the genome median problem, the SPP has been proved to be NP-hard for most genome rearrangement models \cite{PeerS98,Caprara03,TannierZS09,jcb/Kovac14}. 
The only strong tractability result for the SPP  has been obtained in the \emph{Single-Cut-or-Join} (SCJ) model~\cite{FeijaoM11,LuhmannLT17}.

Most methods discussed above consider gene orders with no duplicated genes. 
However gene duplication and loss play an important role in genome evolution.
Unfortunately, in most cases, even computing the pairwise distance between genomes with duplicates is hard~\cite{tcbb/BlinCFRV07,jgaa/AngibaudFRTV09} and there are very few exact polynomial-time algorithms for reconstructing ancestral gene orders in a framework including gene duplication and loss. 
The first work toward this goal was due to Sankoff and ElMabrouk~\cite{Sankoff2000} (see also~\cite{Chauve2013}), who introduced the idea of using reconciled gene trees to define the gene content of ancestral genomes and orthology relations between genes.
This idea was later used in the homology-based method DUPCAR~\cite{MaRR08} that requires however a dated species tree to order gene duplication events, and the DeCoSTAR method that does not require dated gene trees~\cite{DucheminAP17,ANOPHELESc}. 
Other methods accounting for gene duplication and loss include~\cite{Earnest-DeYoungLM2004}, GapAdj~\cite{GagnonBE12} that assumes that gene duplications originate from Whole-Genome Duplications (WGD), PMAG++~\cite{journal.pone.0108796,ZhouT2015}, that relies on a binary encoding of adjacencies and a probabilistic evolutionary model of binary characters to infer ancestral adjacencies, and the homology-based method  MULTIRES~\cite{RajaramanM16} that requires a preliminary set of CARs as well as an upper bound on the number of duplications per gene.
Yet, due to the intractability of the pairwise distance problem in most genome rearrangement models accounting for gene duplication and loss, few methods have been developed to address the SPP.
Recently, it was shown that the problem of computing the pairwise distance in a model including SCJ and single-gene duplications is tractable, although its extension to the  median problem, is intractable~\cite{ManeLFC20}.
Nevertheless, the former result allowed to design a simple \emph{Integer Linear Program} (ILP) for the SPP; however, results on simulated data showed that accounting for gene duplication through single-gene events leads to the evolutionary cost being dominated by duplications and losses, resulting in inaccurate ancestral gene orders~\cite{Mane18}.

Recently, fast ILP-based method were developed for computing the pairwise distance between gene orders with duplicates, especially in genome rearrangement models based on the \emph{Double-Cut and Join} (DCJ) operation~\cite{YancopoulosAF05,BergeronMS06}. 
ILP-based methods for the pairwise distance with duplicated genes can be traced back to the work of Shao~\etal~\cite{ShaoLM15a}, although it was limited to \textit{balanced gene orders}, where both considered genomes are assumed to have the same gene content.
The most recent advance is due to Bohnenk\"amper~\etal~\cite{BohnenkamperBDS20}, who designed an extremely efficient ILP for computing the pairwise distance in the \textit{DCJ-indel} model, that allows arbitrary gene orders and considers DCJ for genome rearrangement operations, and the insertion and deletion of segments of consecutive genes for duplications and losses.
We refer to~\cite{BohnenkamperBDS20} for a more detailed review of previous works. 
In our work, we extend the method introduced in~\cite{BohnenkamperBDS20} to the SPP. 
We also formulate the problem of linearizing degenerate genomes that connects to that of linearizing ancestral circular chromosomes~\cite{Avdeyev2019,Avdeyev2020}. 

We first describe the general workflow of ancestral gene order reconstruction and background on the DCJ-indel distance (Section~\ref{sec:background}), followed by our SPP algorithm (Section~\ref{sec:methods}) before providing results on simulated data (Section~\ref{sec:results}), that show that our algorithm is able to recover efficiently and accurately ancestral gene orders even in the presence of a high level of noise in the input data.

\section{Background}\label{sec:background}

\subsection{Overview}
The parsimony-based ancestral reconstruction approach to which our work contributes aims at inferring gene order sequences for internal nodes of a given rooted species phylogeny based on extant gene orders placed at its tips. 

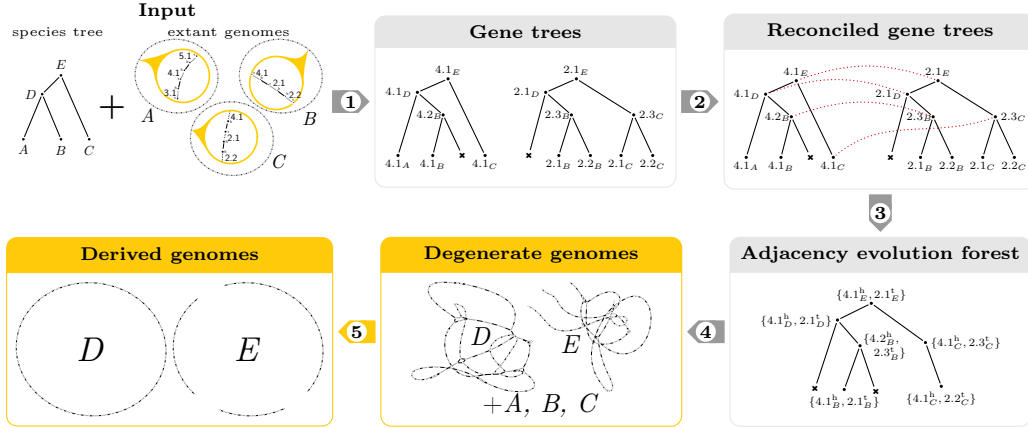
\begin{figure}[tp]
        \hspace{-1cm}\begin{tikzpicture}[block/.style={rectangle split, draw=mygray,
            rectangle split parts=2, text centered, rounded corners, minimum
            height=3cm, minimum width=4cm, outer sep=2pt, font=\scriptsize},
            arrow/.style={-Triangle Cap, line width=2ex, draw=gray!80},
            inc/.style={rectangle, anchor=center, align=center, outer sep=0,
            inner sep=0 },
            wstep/.style={circle, fill=white, midway, inner sep=0.1ex, 
            outer sep=0, font=\scriptsize\bfseries}]
            \matrix[matrix of nodes,column sep=0,nodes=inc] (input) {
                \tiny{species tree} & & \tiny{extant genomes}\\
                \includetikzgraphics[speciestree]{main-diagrams} &
                \hspace{-1ex}{\large \boldmath$+$} & 
                \includegraphics[width=2.5cm]{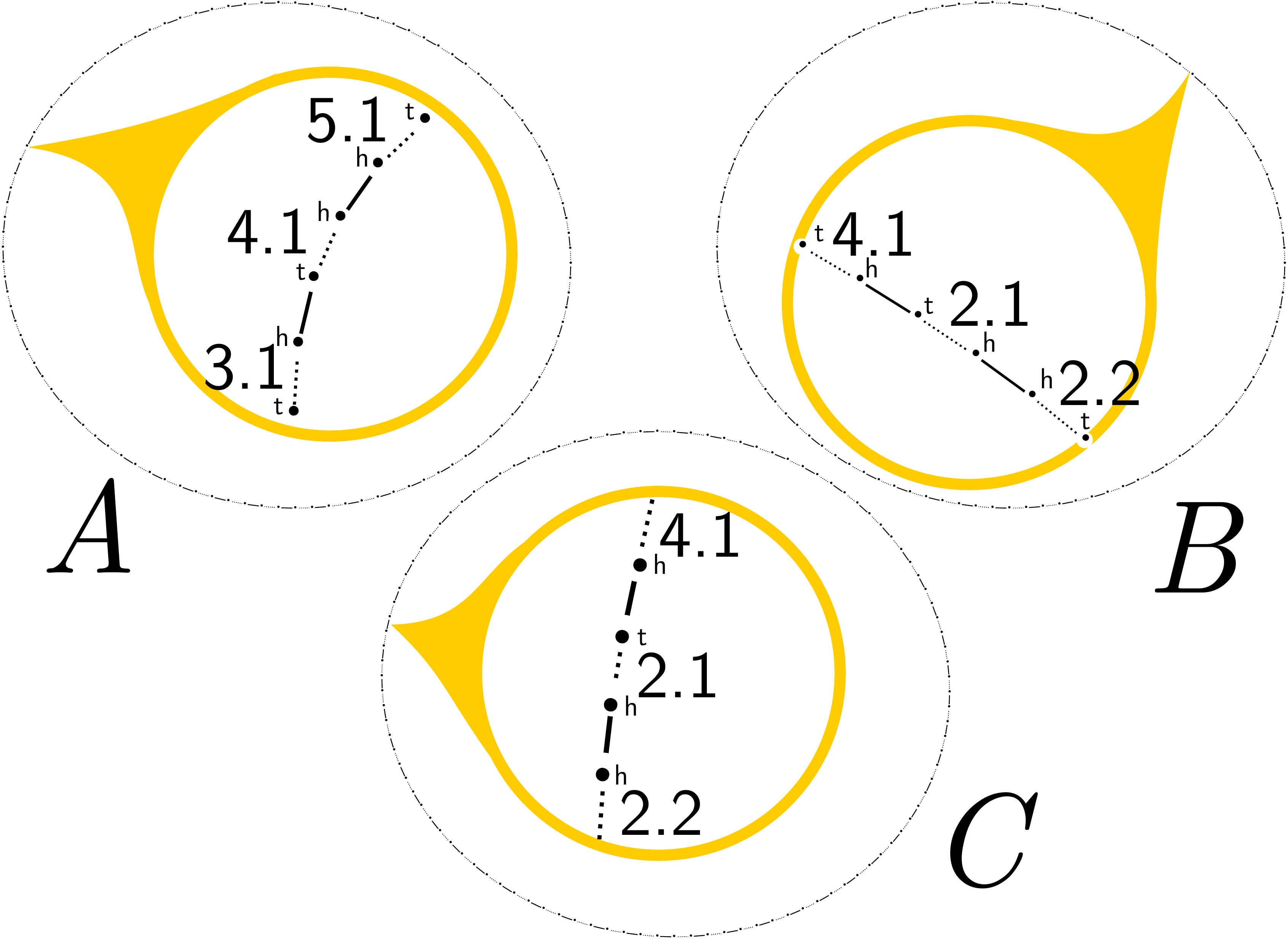}\\
            };

        \node [label={[label distance=-1ex]90:\scriptsize{\textbf{Input}}},
        fit=(input), outer sep=0, inner sep=0] (w1) {};
        \node [block, rectangle split part fill={mygray,white},right=0.5cm
        of w1.east] (w2)
        {\textbf{Gene trees }
            \nodepart{two}
            \includetikzgraphics[genetree1]{main-diagrams}\hspace{0.1cm}
            \includetikzgraphics[genetree2]{main-diagrams}
        }; 
        \node [block, rectangle split part fill={mygray,white},right=0.5cm
        of w2.east] (w3)
        {\textbf{Reconciled gene trees}
            \nodepart{two}
            \includetikzgraphics[reconciledtrees]{main-diagrams}
        }; 
        \node [block, rectangle split part fill={mygray,white},below=0.5cm
        of w3.south] (w4)
        {\textbf{Adjacency evolution forest}

            \nodepart{two}
            \includetikzgraphics[adjforest]{main-diagrams}
        }; 
        \node [block, draw=myyellow, rectangle split part
        fill={myyellow,white},left=0.5cm of w4.west] (w5)
        {\textbf{Degenerate genomes}

            \nodepart{two}
            \includegraphics[width=3.2cm]{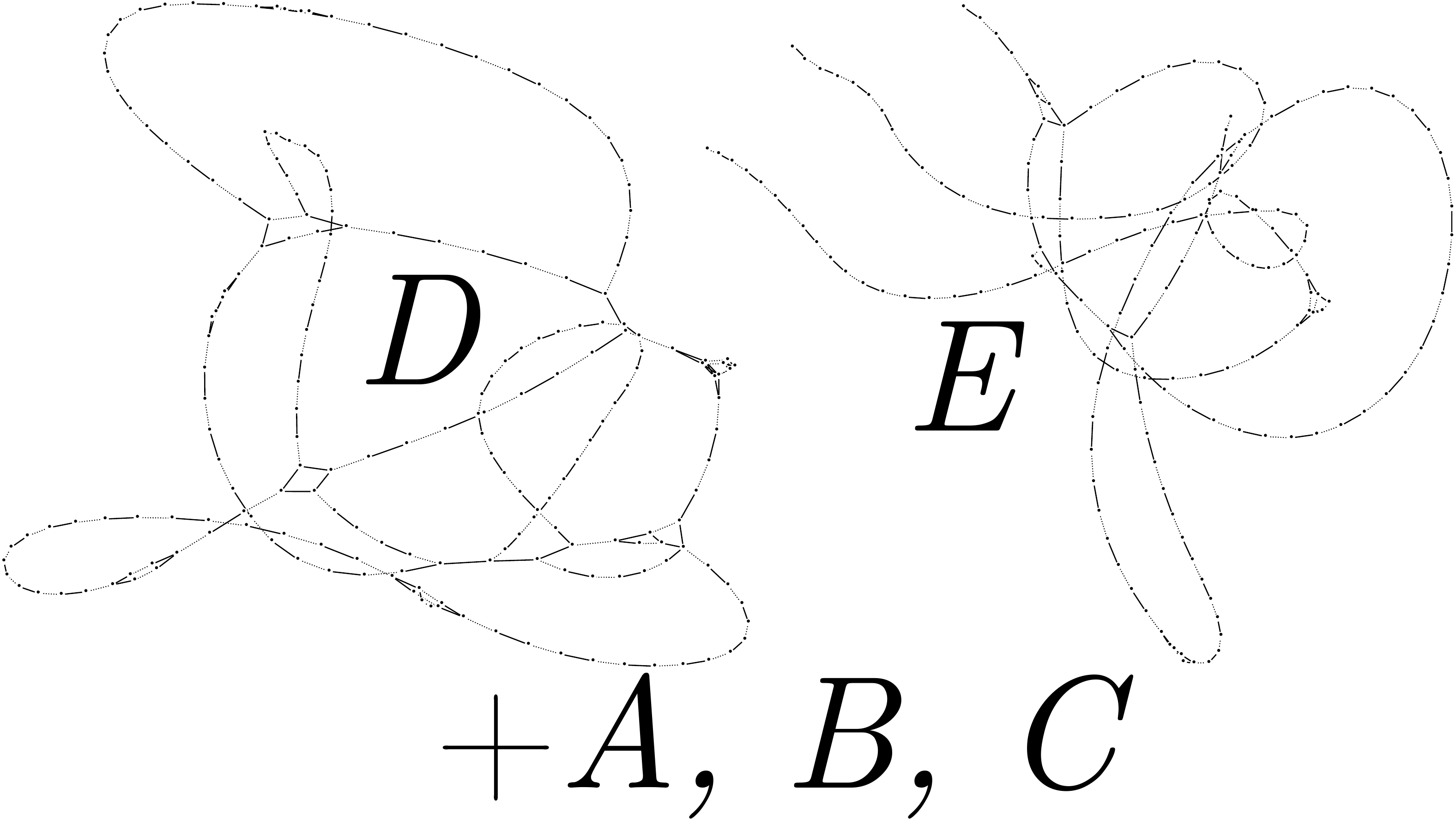}
        }; 
        \node [block, draw=myyellow, rectangle split part
        fill={myyellow,white},left=0.5cm of w5.west] (w6)
        {\textbf{Derived genomes}
            \nodepart{two}
            \includegraphics[width=4.1cm]{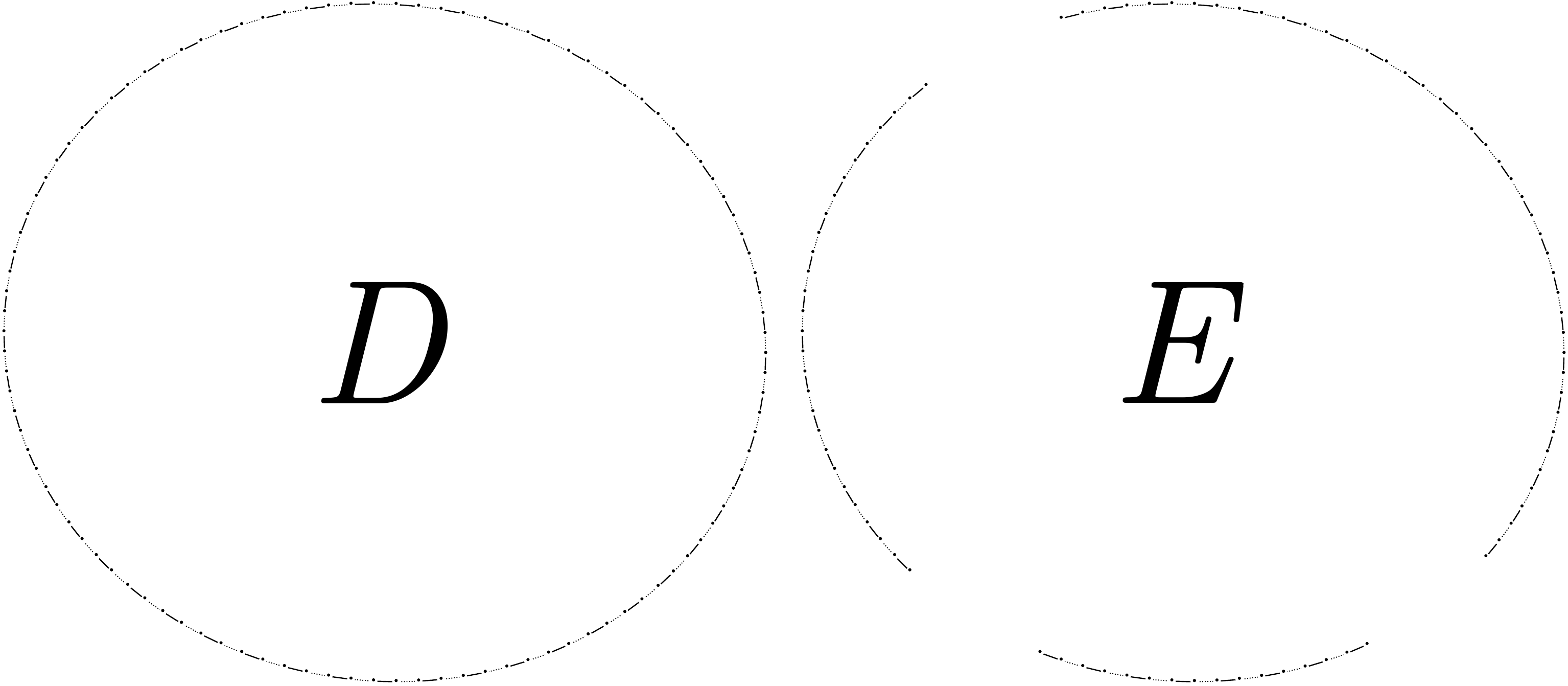}
        }; 

        \draw[arrow] (w1) -- node[wstep]{1} (w2);
        \draw[arrow] (w2) -- node[wstep]{2} (w3);
        \draw[arrow] (w3) -- node[wstep]{3} (w4);
        \draw[arrow] (w4) -- node[wstep]{4} (w5);
        \draw[arrow,draw=myyellow] (w5) -- node[wstep]{5} (w6);
        \end{tikzpicture}
    \caption{
       Ancestral genome reconstruction workflow followed in this paper, from a given phylogeny and genomes at its extant nodes.  The first digit in the label of an illustrated genomic marker, \emph{e.g.} $\mathbf{4}.1$, indicate its gene family membership.
    }\label{fig:workflow}
\end{figure}

From now on, we will refer to gene orders of a species as its \emph{genome}.
In this work, we follow the ancestral genome reconstruction framework that was described in~\cite{Chauve2013} and consists in five successive steps, as illustrated in Figure~\ref{fig:workflow}: 
\begin{enumerate}
    \item Using the underlying DNA or protein sequences of the genomic markers, gene trees are predicted; 
    \item These are subsequently reconciled with the species tree. Note that at this point of the workflow, the marker content of the ancestral genomes subject to reconstruction is determined; 
    \item Reconciled gene trees are then used to infer the evolutionary history of neighborhood relations of adjacent markers, also known as \emph{adjacencies}. 
    In doing so, a phylogenetic model is assumed under which marker adjacencies evolve along an \emph{adjacency tree} through gains, losses, speciations, and lateral gene transfer. 
    While for the first two steps different approaches and a broad variety of software are available, here we rely on the family of DeCo methods that presume parsimonious evolution of adjacencies.
    The latest release, DeCoSTAR~\cite{DucheminAP17}, provides, next to the adjacency evolution forest, weights that are derived from a Boltzmann-Gibbs probabilistic model. 
    These weights are considered as a measure of confidence in ancestral adjacencies and informative for the subsequent steps of the workflow;
    \item From the adjacency forests, ancestral genomes can be derived. However, adjacency forests may conflict, resulting in some markers of ancestral genomes being involved in two or more contradictory adjacencies~\cite{tannier:hal-02535466}. We call a genome \emph{degenerate} if its chromosomes are linear, circular, or neither. That is, an ancestral degenerate genome represents a superposition of a set of ancestral genome candidates;
    \item As the last step of the reconstruction workflow, ancestral genomes are derived from their degenerate counterparts, based on an optimization criterion that considers jointly parsimony in an evolutionary model and adjacency weights. 
\end{enumerate}

In the following, we present a method that addresses step (5) of the ancestral reconstruction workflow by minimizing the sum of pairwise distances between derived genomes along the edges of the species tree. 
The distance that we consider is a general measure of genome rearrangement that takes DCJ events, segmental duplications, segmental deletions, and segmental insertions into account. 

\subsection{Preliminaries}
The following graph notation will be used throughout the manuscript: an (undirected) graph $G = (V, E)$ is a pair of sets $V$ and $E \subseteq V \times V$, denoted node and edge set, respectively. 
We denote by $V(G)$ and $E(G)$ the node and edge set of graph $G$, respectively. 
A \emph{multigraph} is a graph whose edge set is a \emph{multiset}, thus enabling more than one edge between two nodes. 
A weighted graph is a triple $(V, E, \w)$ with $\w: E \to \mathbb R$ a weight function associated to edges. 

\medskip
A \emph{(genomic) marker} $g := \{g^\et, g^\eh\}$ is an element of the universe
of markers, denoted by $\markers$, defined as a pair of \emph{extremities} $g^\et$
(``\emph{tail} of $g$'') and $g^\eh$ (``\emph{head} of $g$''); markers corresponds to genome segments (genes, synteny blocks, \dots).
In what follows we assume that in a gene order, defined as a total order on marker extremities, the two extremities of a marker are always consecutive, \textit{i.e.}, markers do not overlap.
Assuming the doubled stranded nature of DNA, the order of the extremities of a marker encode the strand in which a marker is located: if the tail occurs before the head the marker is assumed to be on the positive strand, and on the negative strand otherwise.
\emph{Telomeres} $\telomeres \subset \markers$ form a special subset of markers composed of a single extremity, \emph{i.e.}, $t := \{t^\circ\}$ for all $t \in \telomeres$; intuitively  both the tail and the head are confounded and  telomeres are not associated to a specific strand.

Let $\bigcup$ be the operator that takes the union of a collection of sets, \emph{i.e.}, $\bigcup {\cal X} := \cup_{X \in {\cal X}} ~X$. 
We denote the universe of (marker) extremities $\bigcup \markers$ by $\extremities$.
Furthermore, we use a function $\extf : \extremities \to \{\et, \eh, \circ\}$ to map extremities to their corresponding kind (tail, head or telomere). 

An \emph{adjacency} is an unordered pair of extremities $\{\varepsilon,
\varepsilon'\} \in \extremities \times \extremities$ such that $\varepsilon\neq
\varepsilon'$. A \emph{genome} $A$ is a set of unique\footnote{No two genomes contain
the same marker and hence the same adjacency.} adjacencies for which
holds true that 
    \emph{(i)} $\forall g^\et \in \bigcup
        A$, there exists also extremity $g^\eh \in \bigcup A$ and vice versa, and
    \emph{(ii)} each extremity is used only once, \emph{i.e.},  $\forall \{X,
        X'\} \subseteq A$, $X \cap X' = \emptyset$.
As a consequence of \emph{(ii)}, it is implicit that the markers of a genome can be ordered in linear and circular segments where the tail and head of any marker are consecutive. 

Comparing genomes benefits from the knowledge of evolutionary relationships between non-telomeric markers. 
Typically, non-telomeric markers are clustered into \emph{families} based on \emph{homology} or \emph{orthology}, indicating a likely common evolutionary origin from an ancestral marker. 
We model family assignments of marker extremities as a function $\family: \extremities\setminus\telomeres \to \mathbb N$ for which holds true that for any marker $g = \{g^\et, g^\eh\}$, $\family(g^\et) = \family(g^\eh)$.  
Function 
$\multiplicity_A(\varepsilon) := |\{\varepsilon' \in \bigcup A \mid \family(\varepsilon) = \family(\varepsilon') \mbox{ and } \extf(\varepsilon) = \extf(\varepsilon')\}|$ 
indicates the \emph{multiplicity} of an extremity's family in a given genome $A$, defined as the number of extremities of same kind that belong to the same family in $A$. 
A family assignment $\family'$ is \emph{$\family$-derived} if there exists no two extremities $\varepsilon, \varepsilon'$ with $\family'(\varepsilon) = \family'(\varepsilon')$ and $\family(\varepsilon) \neq \family(\varepsilon')$.  
A family assignment $\family$ is termed \emph{$\mathcal A$-resolved} (or simply \emph{resolved} if the context is clear) if for each member $A$ of a set of genomes $\mathcal A$, each extremity $\varepsilon \in \extremities \setminus \telomeres$ has a multiplicity not higher than $1$, \emph{i.e.}, $\multiplicity_A(\varepsilon)\leq 1$.  

\subsection{Genome rearrangement model}
A \emph{Double-Cut-and-Join} (DCJ) operation produces a new genome $A'$ from a given genome $A$ by either acting on a single adjacency or a pair of adjacencies. 
In the former case, telomeric adjacencies are created by ``splitting up'' a given adjacency $X = \{a_1, a_2\}$, $X \in A$ and replacing it with a new pair of telomeric adjacencies, \emph{i.e.}, $A' = A \setminus \{X\} \cup \{\{a_1, t\}, \{a_2, t'\}\}$, with $\{t, t'\} \subseteq \telomeres$. 
In the latter case, the DCJ operation acts on a given a pair of adjacencies $X = \{a_1, a_2\}$, $Y = \{b_1, b_2\}$, $\{X, Y\} \subseteq A$, as follows:
\begin{itemize}
    \item $A' = A \setminus \{X, Y\} \cup \{\{a_1, b_1\}, \{a_2, b_2\}\}$, or
    \item $A' = A \setminus \{X, Y\} \cup \{\{a_1, b_2\}, \{a_2, b_1\}\}$, or
    \item $\{a_1, b_1\} \subseteq \telomeres \Rightarrow A' = A \setminus \{X,
        Y\} \cup \{\{a_2, b_2\}\}$ (telomeres $\{a_1, b_1\}$ are removed).
\end{itemize}

An \emph{indel} operation either inserts or deletes a continuous segment of non-telomeric markers of a chromosome, or an entire linear or circular chromosome. 
The unrestricted use of indels results in solutions to the rearrangement problem, where indels of entire chromosomes dominate the rearrangement scenario. 
Such biologically irrelevant solutions are prohibited by restricting the removal and insertion of non-telomeric markers to the minimal number that is needed so that the multiplicity of each marker family in one genome equals that of the other. 
In other words, we do not allow any intermediate genome to have less markers of a family than the minimum of the multiplicity of this marker in both input genomes.
This restriction, also called \emph{maximum matching model}, prevails for the remainder of this paper.

\subsection{Distance calculation} 
For two genomes $A, B$ and an $\{A, B\}$-resolved family assignment $\family$, the \emph{DCJ indel distance} is defined as $\dist_\DCJid(A, B) := d$, were $d$ is the minimum number of DCJ/indel operations $\Delta_i$ in any rearrangement scenario $A \xrightarrow{\Delta_1} A_1 \xrightarrow{\Delta_2} A_2 \cdots A_{d-1}\xrightarrow{\Delta_d} B $ that transforms $A$ into $B$. 

\begin{problem}[DCJ indel distance~\cite{Braga:2011kz}]\label{prb:dcj_indel}
    Given two genomes $A, B$ and an $\{A,B\}$-resolved family assignment
    $\family$, calculate the DCJ indel distance $\dist_\DCJid(A, B)$. 
\end{problem}

In the following, we briefly outline a solution to Problem~\ref{prb:dcj_indel} introduced in~\cite{Braga:2011kz}.
It is convenient to make use of a graph structure to calculate the DCJ-indel distance.  
The \emph{relational diagram} $R(A, B, \family)$ of two genomes $A$ and $B$ and $\{A, B\}$-resolved family assignment $\family$ is a multigraph $G = (V^{AB}, E^{AB})$ with nodes $V^{AB} = V^A \cup V^B$ representing marker extremities of genomes $A$ and $B$. 
The graph $G$ has three types of edges: (i) \emph{adjacency edges} $E_\adj^{AB} = E_\adj^A \cup E_\adj^B$, corresponding to the adjacencies of $A$ and $B$, (ii) \emph{extremity edges} $E_\ext^{AB}$, that connect marker extremities between the two genomes according to $\{A, B\}$-resolved family assignment $\family$, and (iii) \emph{indel edges} $E_\id^{AB} = E_\id^A \cup E_\id^B$, each of which connects the extremities of a marker that is removed or inserted in the respective genome.  

\begin{figure}[tb]
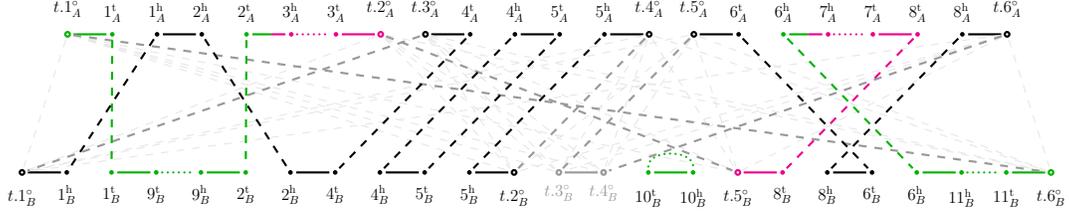

    \hspace{-1cm}\includetikzgraphics[relational_diagram]{main-diagrams}
    \caption{
        \textbf{Non-gray part}: Relational diagram between two exemplary genomes $A$ and $B$. 
        Solid lines show edges of $E_\adj^{AB}$, dotted lines show edges of $E_\id^{AB}$, and dashed lines show edges of $E_\ext^{AB}$. 
        Paths of $A$- and $B$-runs (dotted lines) are highlighted in green and magenta, respectively. 
        \textbf{Dark-gray part:} Additional telomeric extremities and telomeric extremity edges that, together with black and colored edges make up a capped relational diagram of $A$ and $B$. 
        \textbf{Light-gray edges} correspond to candidates of telomeric adjacencies.
    }\label{fig:rd}
\end{figure}
\begin{example}
    The non-gray part of Figure~\ref{fig:rd} depicts the relational diagram of two exemplary genomes, $A$ and $B$.
\end{example}

Note that under the maximum matching model, the $\{A, B\}$-resolved family assignment $\family$ dictates which marker is removed or inserted, that is, $E_\id^A = \{ \{g^\et, g^\eh\} \subseteq \bigcup A \mid \multiplicity_B(g^\et) < 1\}$, and $E_\id^B = \{ \{g^\et, g^\eh\} \subseteq \bigcup B \mid \multiplicity_A(g^\et) < 1\}$.  
Subsequently, we denote by $n$ the number of non-telomeric markers of one genome that are not inserted or deleted, which is identical for both genomes $A$ and $B$:
$$
   n := \frac{1}{2}|\{\varepsilon \in \bigcup A \setminus \telomeres \mid
   \multiplicity_B(\varepsilon) = 1\}| \quad = \frac{1}{2}|\{\varepsilon \in
   \bigcup B \setminus \telomeres \mid \multiplicity_A(\varepsilon) = 1\}|
$$

Each node in $V^{AB}$ has degree one or two, so the connected components of the graph constitute alternating paths and cycles and each node corresponding to a non-telomeric marker extremity is incident to one adjacency edge and either one extremity or one indel edge. 
Telomeric extremity nodes have degree one and are incident to an adjacency edge. 

The DCJ-indel distance, given by $\dist_\DCJid(A, B) = n - c - \frac{i}{2} + \delta$, grows inversely to the number $c$ of cycles and twice the number $i$ of paths of odd length in $G$~\cite{BergeronMS06,Braga:2011kz}, where $\delta$ denotes the \emph{indel penalty}, that is, the number of indel operations required to transform $A$ into $B$.
The maximal contribution of a connected component of the relational diagram to the \emph{indel penalty}, the \emph{indel potential}, is quantified based on the concept of runs: an \emph{$A$-run} is a maximal sequence of indel edges of genome $A$ $(e_1, \ldots, e_l)$, with $e_1, \ldots, e_l \in E_\id^A$, that lies on a connected component of graph $G$, that does not pass over an indel edge of genome $B$. 
Analogously, a $B$-run is a maximal sequence of indel edges of genome $B$ that does not pass over an indel edge of genome $A$. 
For a given connected component $C \in G$, let $\mathbf{\Lambda}(C)$ denote the number of runs it contains; the \emph{indel-potential} of $C$ is then given by
$$
    \mathbf{\lambda}(C) = 
    \begin{cases} 
        0                                                        & \mbox{if
        $\mathbf{\Lambda}(C)=0$ and} \\
        \left\lceil \frac{\mathbf{\Lambda}(C)+1}{2}\right\rceil  & \mbox{otherwise.}
    \end{cases}
$$
The sum of indel potentials of all connected components of graph $G$ serves as an upper bound of indel penalty $\delta$: 
$\delta \leq \sum_{C \in G} \mathbf{\lambda}(C)$.

The calculation of the \emph{exact} indel penalty must take into account further DCJ-induced recombinations of connected components that reduce the number of necessary indel operations by merging successive runs. 
Braga \etal~ show that only recombinations of those connected components that are \emph{paths} lead to optimal rearrangement scenarios. 
There are $32$ groups of path recombinations that can be further classified into five categories with varying negative contributions to the indel potential of its recombinants.
Further details can be found in~\cite{Braga:2011kz}. 
With the help of tabulation, the indel penalty---and thus the DCJ indel distance---of two genomes under a resolved family assignment can be calculated in linear time~\cite{Braga:2011kz}. 

\smallskip
Bohnenk\"amper \etal~describe an alternative approach to calculate the indel penalty that makes use of a technique called \emph{capping}~\cite{ShaoM17}, where paths are assembled into alternating cycles that correspond to optimal path recombinations. 
This requires the extension of the relational diagram to a \emph{capped multi-relational diagram} $\MR(A, B, \family)$~\cite{BohnenkamperBDS20} in which node degree is no longer restricted to one  and two. 
In fact, in a capped multi-relational diagram, each node has degree two or higher. 

To calculate the indel penalty in the above-described setting, we construct the capped multi-relational diagram $H$ from the existing relational diagram $G$, i.e, $H \gets G$.
Then, if genomes $A$ and $B$ have unequal numbers of telomeres, additional nodes and edges corresponding to \emph{telomeric adjacencies} are added to the graph and are associated with the genome that has fewer telomeres.
Without loss of generality, let $A$ be the genome with the lower number of telomeres and $\tau := |\bigcup B\cap \telomeres| - |\bigcup A \cap \telomeres|$. 
Note that as a result of the genome constraints, $\tau$ is even.
Then $\tau$ new telomeric extremities $\{\varepsilon_{1}, \ldots, \varepsilon_{\tau}\} \subseteq \telomeres$ are added to $V_A(H)$. 
The newly added nodes are pairwise connected by adjacency edges $\{\varepsilon_{2i-1}, \varepsilon_{2i}\}$, $1 \leq i \leq \frac{\tau}{2}$, and associated with edge set $E_\adj^A(H)$. 
Second, additional extremity edges are added to edge set $E_\ext^{AB}(H)$, corresponding to the Cartesian product of telomeric extremities $(\bigcup A \cap \telomeres) \times (\bigcup B \cap \telomeres) =: T$. 
The indel penalty $\delta$ is determined by finding a subset of extremity edges $T' \subset T$ in $H$ corresponding to a perfect matching between telomeric nodes $V^A(H)\cap \telomeres$ and $V^B(H) \cap \telomeres$  such that the sum of indel potentials of all connected components of the graph is minimized. 
Observe that any perfect matching $T'\subset T$ decomposes the graph into
alternating cycles, thereby transforming $H$ into a \emph{capped} relational
diagram $R_\circ(A, B, \family)$.

If such constructed capped relational diagram $G_\circ$ is given, the DCJ-indel
distance can be expressed in terms of connected components and their indel
potentials:
\begin{align}\label{eq:dcj_id}
    \dist_\DCJid(A, B) = n' - \sum_{C \in G_\circ} 1-\mathbb{1}_{E^{AB}_\ext(C) =
    \emptyset}- \lambda(C)\,,
\end{align}
where $\mathbb{1}$ denotes the indicator function and
$n' = n + \frac{V^{AB}(H) \cap \telomeres}{4}\,$.

Cycles that do not contain any extremity edge\footnote{Note that the number of extremity edges in a cycle of a relational diagram is even.} correspond to circular chromosomes that are entirely inserted or deleted in the rearrangement scenario between the two genomes. 
These connected components of the relational diagram are termed \emph{circular singletons} and differ from extremity-edge enclosing cycles in that their presence does not contribute to the count of cycles that reduces the distance but only increases the indel penalty.

Bohnenk\"amper~\etal indirectly quantify the indel potential by counting
transitions between runs of a connected component: For a given connected
component $C$ of the capped relational diagram, a \emph{transition} is a path
$\{\varepsilon_1, \varepsilon_2\}, \{\varepsilon_2, \varepsilon_3\},
\allowbreak \ldots, \{\varepsilon_{l-1}, \varepsilon_l\} \in E_\adj^{AB}(C) \cup E_\ext^{AB}(C)$ such that their adjacent edges are indel edges of two different runs, \emph{i.e.}, without loss of generality $\{\varepsilon_0, \varepsilon_1\} \in E_\id^{A}$ and $\{\varepsilon_l, \varepsilon_{l+1}\} \in E_\id^{B}(C)$. 
To count transitions, one of the edges of the transition is designated as \emph{transition edge}, which we here arbitrarily define as edge $\{\varepsilon_1, \varepsilon_2\}$. 
Note that by construction, $\{\varepsilon_1, \varepsilon_2\}$ must be part of the edge set $E_\adj^A$. 
With that, the DCJ-indel distance can be separated into four terms 
\begin{align*}
    \dist_\DCJid(A, B) = & n' - \displaystyle\sum_{C \in G_\circ}
    \mathbb{1}_{E_\id^{AB}(C) = \emptyset} + \frac{1}{2}
    \displaystyle\sum_{e \in E^{A}_\adj(G_\circ)} \mathbb{1}_{\text{$e$ is transition
    edge}} \\
    & + \sum_{C \in G_\circ} \mathbb{1}_{E_\ext^{AB}(C)=\emptyset}\,.
\end{align*}

\addtocounter{example}{-1}
\begin{example}[cont'd]
    Figure~\ref{fig:rd} shows the capped multi-relational relational diagram of
    $A$ and $B$.  Black, colored, and dark-gray parts indicate the capped
    relational diagram of a solution to Problem~\ref{prb:dcj_indel}. The resulting DCJ-indel distance is $d_\DCJid(A, B) = 9 - 4 + \frac{1}{2} \cdot 2 + 1 = 7$.
\end{example}

\section{Methods}\label{sec:methods}

\subsection{The small parsimony problem for degenerate genomes}
A \emph{degenerate genome} is a set of unique adjacencies $A$ that satisfies the following conditions:
    \emph{(i)} $\forall g^\et \in \bigcup A$, there exists also extremity $g^\eh \in \bigcup A$ and vice versa, and 
    \emph{(ii)} each telomeric extremity is used only once: $\forall \{X, X'\} \subseteq A$, $X \cap X' \cap \telomeres = \emptyset$. 
Note that a genome is also a degenerate genome, but the reverse does not hold true in general. The \emph{surfeit} of a degenerate genome $A$ is the ratio $\frac{2\cdot|A|}{|\bigcup A \setminus \telomeres|}\,$.

\begin{figure}[tb]
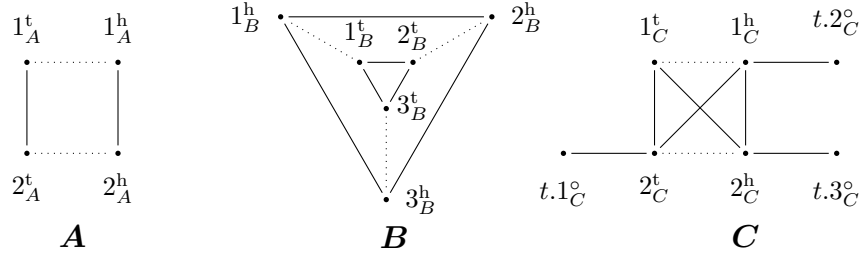

    {\centering
    \includetikzgraphics[degenerate_genomes]{main-diagrams}\\}
    \caption{Visualization of genome $A$, degenerate genome $B$, and linearizable degenerate genome $C$. 
    Solid edges correspond to adjacencies, dotted lines connect extremities of the same marker.
    }
    \label{fig:degenerate_genomes}
\end{figure}

\begin{example}\label{ex:degenerate_genomes}
    Consider adjacency sets $A = \{ \{1_A^\et, 2_A^\et\}, \{1_A^\eh,
    2_A^\eh\}\}$, $B = \{ \{1_B^\eh, 2_B^\eh\},\allowbreak \{1_B^\eh, 3_B^\eh \},
    \allowbreak \{2_B^\eh, 3_B^\eh\}, \{1_B^\et, 2_B^\et\}, \{1_B^\et,
    3_B^\et\}, \{2_B^\et, 3_B^\et\}\}$, $C = \{ \{1_C^\et, 2_C^\et\},
    \{1_C^\eh, 2_C^\eh\}, \{1_C^\et, \allowbreak 2_C^\eh\},\{1_C^\eh, 2_C^\et\},
    \{t.1_C^\circ,\allowbreak  2_C^\et\}, \{t.2_C^\circ, 1_C^\eh\},
    \{t.3_C^\circ, 2_C^\eh\}\}\}$,
    as illustrated in Figure~\ref{fig:degenerate_genomes}. All three are degenerate genomes, but only $A$ is a genome. Their surfeits are $1.0$ and $2.0$, and $3.5$, respectively.
\end{example}

A genome $A'$ is \emph{$A$-derived} from degenerate genome $A$ (or simply ``derived from $A$'') if $A' \subseteq A$ and $\bigcup A' \setminus \telomeres= \bigcup A \setminus \telomeres$. 
Conversely, a degenerate genome $A$ is \emph{linearizable} if there exists an $A$-derived genome. 
In fact, many degenerate genomes are not linearizable. 

\addtocounter{example}{-1}
\begin{example}[cont'd]
    Degenerate genome $B$ is not linearizable. Degenerate genome $C$ is linearizable, and $\{ \{1_C^\et, 2_C^\et\}, \{1_C^\eh, 2_C^\eh\} \}$, $\{ \{1_C^\et, \allowbreak 2_C^\eh\}, \{1_C^\et, 2_C^\eh\} \}$, $\{\{1_C^\et, 2_C^\et\}, \{t.2_C^\circ, 1_C^\eh\}, \allowbreak \{t.3_C^\circ, 2_C^\eh\}\}$, $\{\{1_C^\et, 2_C^\eh\}, \{t.1_C^\circ, 2_C^\et\}, \allowbreak \{t.2_C^\circ, 1_C^\eh\}\}$ are all $C$-derived genomes.
\end{example}

\subsection{Ensuring linearizability of degenerate genomes}
The ancestral adjacencies that DeCoSTAR infers in the third step of the ancestral reconstruction workflow do not guarantee that the resulting degenerate genomes are linearizable. 
Whether a polynomial algorithm exists that can test whether a given degenerate genome $A$ is linearizable is an open question. 
Therefore, to ensure linearizability, we augment degenerate genomes with additional telomeric adjacencies. 
That is, for a non-telomeric extremity $\varepsilon \in \bigcup A\cap \telomeres$ of a degenerate genome $A$ such that $\varepsilon$ is not already involved in an adjacency with a telomeric extremity, 
we add a new telomeric adjacency $\{\varepsilon, t_\varepsilon\}$, $t_\varepsilon \in \telomeres$, to $A$ and assign it weight $0$.
From this procedure parts of the degenerate genome are omitted that are linearizable. 
To this end, we study connected components of the graph $G_A = (V, E)$ where $V = \bigcup A$ and $E=A$. 
Nodes of a connected component $C \in G_A$ do not need to be augmented with telomeric adjacencies if $|C|$ is even and $C$ is a cycle, a path, or fully connected. 

\subsection{The small parsimony problem for degenerate genomes}
In the following, we study a variant of Problem~\ref{prb:dcj_indel} in which linearizable degenerate genomes and an unresolved family assignment are given with the objective to derive genomes and a resolved family assignment that minimize the DCJ indel distance.  
For a given pair of linearizable degenerate genomes, often multiple solutions to the DCJ indel distance problem exist.
That is, different choices of derived genomes, resolved family assignments, as well as rearrangement scenarios lead to the same optimal DCJ indel distance. 
The solution space increases with the surfeit of the degenerate genomes as well as the multiplicity of marker extremities under the given unresolved family assignment.
At the same time, solutions based on derived genomes that contain excessive numbers of linear chromosomes are not likely to occur in the biological realm. 
To reduce the solution space as well as to ameliorate biological interpretability, we aim to find resolved genomes with fewer linear chromosomes even if that comes at the expense of an increased DCJ indel distance. 
At last, we want to facilitate the consideration of prior knowledge about presence or absence of certain adjacencies in degenerate genomes, if such is available:

\begin{problem}[Weighted Degenerate DCJ indel distance]\label{prb:wdeg_dcj}
    Given a weighting scheme $\w : \extremities \times \extremities \to \mathbb R$, some $\alpha,\beta \in [0, 1]$, two linearizable degenerate genomes $A, B$ and family assignment $\family$, find $A$-derived genome $A'$, $B$-derived genome $B'$, and $\family$-derived $\{A',B'\}$-resolved family assignment $\family'$ that minimize the linear combination
    $$
    (1-\alpha-\beta) \cdot \sum_{X \in A' \cup B'} -\w(X) + \alpha \cdot \dist_\DCJid(A', B') + \beta \cdot |(\bigcup A' \cup \bigcup B')\cap \telomeres|\,.
    $$
\end{problem}

Because the surfeit of a derived genome only depends on the number of its telomeres, minimizing its number of telomeres is equivalent to minimizing its  surfeit. 

In this work, we study the following generalization of Problem~\ref{prb:wdeg_dcj} that represents a variant of the \emph{small parsimony problem} (SPP) under the weighted degenerate DCJ indel distance. 
To this end, we make use of an additional graph that establishes relationships between a given set of genomes $\mathcal A = A_1, \ldots, A_k$ and that we call ``phylogeny''. 
A \emph{phylogeny} $\phylo$ is a connected graph with nodes  $V(\phylo) = \mathcal A$. 
Nodes $L \subseteq V(\phylo)$ of degree 1 are termed \emph{leaves}.

\begin{problem}[SPP-DCJ]\label{prb:spp_dcj}
    Given a phylogeny $\phylo$ and a set of linearizable degenerate genomes $A_1, \ldots, A_k$ corresponding to the node set $V(\phylo) = \{A_1, \ldots, A_k\}$, find genomes $A'_1 \subseteq A_1, \ldots, A'_k \allowbreak \subseteq A_k$ that minimize the sum of weighted degenerate DCJ indel distances along the edges of $\phylo$. 
\end{problem}

\subsection{Finding solutions to SPP-DCJ}
Our solution to SPP-DCJ makes use of the capped multi-relational diagram as introduced in~\cite{BohnenkamperBDS20}, subsequently simply denoted by \emph{multi-relational diagram}. 
The multi-relational diagram represents a superposition of capped relational diagrams of all possible derived genomes of degenerate genomes $A$ and $B$. 
For each edge $\{A, B\}$ of a given phylogeny $\phylo$ and a given family assignment $\family$, we construct multi-relational diagram $(V^{AB}, E^{AB}) = \MR(A, B, \family)$. 
Similarly to the relational diagram introduced in Section~\ref{sec:background}, the multi-relational diagram is a composition of node sets $V^{AB} = V^A \cup V^B$ and edge sets $E^{AB} = E^{AB}_\adj \cup E^{AB}_\ext \cup E^{AB}_\id$.
\begin{itemize}
    \item A node of the multi-relational diagram corresponds either to an
        extremity of degenerate genomes $A$ or $B$, or to an additional
        telomeric extremity required to match the number telomeres in both
        degenerate genomes. Here, we take advantage of the following
        observation: Although the number of telomeres of a degenerate genome
        can be odd, only an even subset thereof can be simultaneously part of a
        derived genome. In other words, if a degenerate genome has $x$
        telomeres, at most $2\lfloor\frac{x}{2}\rfloor$ telomeres need
        counterparts in the opposing degenerate genome. Let $l = |\bigcup B
        \cap \telomeres | - |\bigcup A \cap \telomeres |$ be the difference of
        telomere counts of both degenerate genomes. Then the set of additional
        telomeres added to the relational diagram of $A$ and $B$ is defined by
        sets $T^A = \{t_i \in \telomeres \mid 1 \leq i \leq 2 \lfloor
        \frac{l}{2} \rfloor \}$ and $T^B = \{t_i \in \telomeres \mid 1 \leq i
        \leq 2 \lfloor \frac{-l}{2} \rfloor \}$.  Observe that by construction,
        the size of both sets is even and at most one of the two sets is
        non-empty. Then the node sets associated with degenerate genomes $A$
        and $B$ are defined by $V^A = \bigcup A \cup T^A$ and $V^B = \bigcup B
        \cup T^B$.
    \item Likewise, the set of adjacency edges $E_\adj^{AB} = E_\adj^A
        \cup E_\adj^B$ comprises adjacencies of $A$ and $B$ as well as
        additional telomeric adjacencies between nodes of $T^A$ or $T^B$, \emph{i.e.},
        $E_\adj^A = A \cup \{\{t_{2i-1}, t_{2i}\} \mid 1 \leq i
        \leq 2 \lfloor \frac{l}{2} \rfloor\}$, $E_\adj^B = B \cup \{\{t_{2i-1},
        t_{2i}\} \mid 1 \leq i \leq 2 \lfloor \frac{-l}{2}
        \rfloor\}$. 
    \item Extremity edges connect marker extremities of degenerate genome $A$
        with those of degenerate genome $B$ according to family assignment
        $\family$, but they also connect all telomeric extremities of $A$ with
        all telomeric extremities of $B$, \emph{i.e.}, $E_\ext^{AB} = \{\{\varepsilon,
        \varepsilon'\} \in V^A \times V^B \mid \varepsilon, \varepsilon' \in
        \telomeres \text{ or } \family(\varepsilon) = \family(\varepsilon')
        \text{ and } \extf(\varepsilon) = \extf(\varepsilon') \}$.
	\item Indel edges $E_\id^{AB} = E_\id^A \cup E_\id^B$ connect extremities
		of markers whose families are overrepresented in the respective genome,
        \emph{i.e.}, $E_\id^A = \{ \{g^\et, g^\eh\} \subseteq \bigcup A \mid
        \multiplicity_A(g^\et) > \multiplicity_B(g^\et) \}$, and $E_\id^B = \{
        \{g^\et, g^\eh\} \subseteq \bigcup B \mid \multiplicity_B(g^\et) >
        \multiplicity_A(g^\et)\}$. 

\end{itemize}


\begin{algorithm}[hp]
    \small{
    \caption{\label{alg:ilp}ILP for solving Problem SPP-DCJ.}
    
    \smallskip
    \noindent\textbf{Objective:}
    
    \phantom{sp}\texttt{Maximize}
    
    \vspace{-5ex}
    {\footnotesize\begin{align*}
        (1-\alpha-\beta)  \displaystyle\sum_{\substack{X \in A \cup B}} \w(X)
    +   \alpha            \left(\displaystyle\sum_{\substack{1\leq i \leq|V^{AB}\setminus \telomeres|,\\AB \in E(\phylo) }} z_i^{AB} -
                              \frac{1}{2}\cdot \displaystyle\sum_{\substack{e \in E^{AB},\\AB \in E(\phylo)}} t_e^{AB}-
                              \displaystyle\sum_{\substack{C \in \cscandidates^{AB}\\AB \in E(\phylo)}} s_C\right)
    -   \beta             \displaystyle\sum_{\substack{v \in V^{AB} \cap \telomeres\\AB \in E(\phylo)}} o_v
    \end{align*}}
    
    \noindent\textbf{For each multi-relational diagram \boldmath$\MR(A, B, \family)$, \boldmath$\{A, B\}\in E(\phylo)$:}\smallskip\\
    \begingroup
        \renewcommand{\arraystretch}{1.2} 
        \setlength{\tabcolsep}{0.5pt}
        \phantom{sp}\begin{tabular}{p{1.5cm}rp{1.5cm}l}
            \multicolumn{4}{l}{\textbf{Constraints:}\smallskip}\\
            \;(\texttt{C.01}) & $\displaystyle o_v$                                     & $= 1$             & $\forall~v \in V^{AB} \setminus \telomeres$ \\
            \;(\texttt{C.02}) & $\displaystyle\sum_{\{u, v\} \in E_\adj^{AB}}\!\!\! x_{uv}$  & $= o_v$           & $\forall~v \in V^{AB}$ \\
                              & $\displaystyle\sum_{\{u,v\}\in E_\id^{AB} \cup E_\ext^{AB} }\!\!\! x_{uv}^{AB}$ & $= o_v$            & $\forall~v \in V^{AB}$ \\
            \;(\texttt{C.03}) & $x_e^{AB}$                                              & $= x_f^{AB}$      & $\forall~e,f \in E_\ext^{AB} \text{ such that $e$ and $f$ are siblings}$\\
            \;(\texttt{C.04}) & $y_j^{AB} + i(1-x_{v_iv_j})$                            & $\geq y_i^{AB}$   & $\forall~ \{v_i, v_j\}\in E_\adj^{AB}\,,$ \\
                              & $y_j^{AB} + i(1-x_{v_iv_j}^{AB})$                       & $\geq y_i^{AB}$   & $\forall~ \{v_i, v_j\}\in E_\id^{AB} \cup E_\ext^{AB}\,,$ \\
            \;(\texttt{C.05}) & $i (1 - x_{v_iv_j}^{AB})$                               & $\geq y_i^{AB}$   & $\forall~\{v_i,v_j\} \in E_\id^{AB}$\\
            \;(\texttt{C.06}) & $i\cdot z_i^{AB}$                                       & $\leq y_i^{AB}$   & $\forall~1 \leq i \leq |V^{AB} \setminus \telomeres|$ \\ 
            \;(\texttt{C.07}) & $1 - x_{uv}^{AB}$                                       & $\geq r_v^{AB}$   & $\forall~\{u, v\}\in E_\id^A\,,$\\
                              & $x_{u'v'}^{AB}$                                         & $\leq r_{v'}^{AB}$& $\forall~\{u', v'\} \in E_\id^B$ \\
            \;(\texttt{C.08})  & $r_v^{AB} - r_u^{AB} - (1-x_{uv}^{AB})$                & $\geq t_{uv}^{AB}$& $\forall~\{u,v\} \in E^{AB}$ \\
            \;(\texttt{C.09}) & $\displaystyle\sum_{e \in E_\adj^{AB}(C)} \!\!\! x_e + \displaystyle\sum_{e \in E_\id^{AB}(C)} \!\!\! x_e^{AB} + 1$ & $\leq s_C$      & $\forall~C \in \mathcal C^{AB}$\smallskip\\
            \;(\texttt{C.10}) & $\displaystyle\sum_{\substack{d \in E_\id^A \\ d \cap e \neq \emptyset}} \!\!\! x_{d}^{AB}- t_{e}^{AB}$ & $\geq 0$      & $\forall~e \in E_\adj^A$\\
                              & $t_e^{AB}$                                              & $= 0$             & $\forall~e \in E_\id^{AB} \cup E_\ext^{AB}$ \\ 
            \;(\texttt{C.11}) & $\displaystyle\sum_{v \in V^{A} \cap \telomeres}\!\!\! o_v -2a_{A}$  & $= 0$         & \\
                              & $\displaystyle\sum_{v \in V^{B} \cap \telomeres}\!\!\! o_v -2a_{B}$  & $= 0$         & \\

            \multicolumn{4}{l}{\textbf{Domains:}\smallskip}\\
            \;(\texttt{D.01}) & $x_e$                                                   & $\in \{0, 1\}$    & $\forall~e \in E_\adj^{AB}$ \\
                              & $x_e^{AB}$                                              & $\in \{0, 1\}$    & $\forall~e \in E_\id^{AB} \cup E_\ext^{AB}$ \\
            \;(\texttt{D.02}) & $0~ \leq y_i^{AB}$                                      & $\leq i$ & $\forall~1 \leq i \leq |V^{AB}|$\\
            \;(\texttt{D.03}) & $z_i^{AB}$                                              & $\in \{0, 1\}$    & $\forall~1 \leq i \leq |V^{AB} \setminus \telomeres|$\\
            \;(\texttt{D.04}) & $r_v^{AB}$                                              & $\in \{0, 1\}$    & $\forall~v \in V^{AB}$ \\
            \;(\texttt{D.05}) & $t_e^{AB} \in \{0, 1\} $                                & $\in \{0, 1\}$    & $\forall~e \in E^{AB}$ \\ 
            \;(\texttt{D.06}) & $o_v$                                                   & $\in \{0, 1\}$    & $\forall~v \in V^{AB}$ \\
            \;(\texttt{D.07}) & $s_C$                                                   & $\in \{0,1\}$     & $\forall~C \in \mathcal C^{AB}$ \\
            \;(\texttt{D.07}) & $a_A,a_B$                                               & $\in \mathbb N$   &  \\
        \end{tabular}
    \endgroup
    }
\end{algorithm}

Our method for solving Problem~\ref{prb:spp_dcj} is an extension of DING~\cite{BohnenkamperBDS20} and is formulated as an ILP (Algorithm~\ref{alg:ilp}). 
The key idea of the ILP is to simultaneously calculate solutions to Problem~\ref{prb:wdeg_dcj} for each pair of genomes $\{A, B\} \in E(\phylo)$ and family assignment $\family$ while maintaining consistency in the selection of adjacencies of each derived genome among all solutions. 
A solution is encoded as capped relational diagram $R_\circ(A', B', \family')$ and represents $A$-derived genome $A'$, $B$-derived genome $B'$ and $\family$-derived $\{A, B\}$-resolved family assignment $\family'$. 

Rather than minimizing the sum of objectives specified in Problem~\ref{prb:wdeg_dcj}, the algorithm solves the inverse maximization problem. 
Note that $n$, \emph{i.e.}, the number of non-telomeric markers that are shared between each pair of derived genomes $A', B'$, is fixed and amounts to
$$ 
    \frac{1}{2} \sum_{\varepsilon \in A}\min\left(\multiplicity_A(\varepsilon),
    \multiplicity_B(\varepsilon)\right) \quad = \frac{1}{2} \sum_{\varepsilon \in
        B}\min\left(\multiplicity_A(\varepsilon),
        \multiplicity_B(\varepsilon)\right)\,
$$
and thus requires no optimization.
Similarly, each quadruple of telomeres\footnote{Telomeres in a capped
relational diagram appear in multiples of four} may increase the number of
cycles by at most 1 while simultaneously increasing $n'$ by 1 (cf.~Eq.~\ref{eq:dcj_id}): the number of telomeres used in a solution does not impact the DCJ-indel distance.  
The objective of Algorithm~\ref{alg:ilp} translates the linear combination of Problem~\ref{prb:spp_dcj} into weighted sums over binary variables $z_i^{AB}$, $t_e^{AB}$, $s_C$, and $o_v$ that count indel-free cycles, transition edges, circular singletons, and telomeric extremities, respectively. 
Here, $\mathcal C^{AB}$ represents the set of all candidates of circular singletons.

\smallskip
The choice of adjacencies across all relational diagrams that make use of the same genome is synchronized by using the same variable for each of its adjacencies (cf.~D.01). 
Conversely, indel and extremity edges depend on the particular pairwise comparison. 
Also, variables involved in identifying/counting indel-free cycles (cf.~D.02, D.03), indels (cf.~D.04), and transition edges (cf.~D.05) are not synchronized, and therefore are optimized independently. Note the capped relational diagram does not permit components with only telomeric extremities. Thus allows us to omit telomeric extremities from the process of counting indel-free cycles.
As alluded to before, the set of extremities in a genome derived from a degenerate genome may vary, leading to different surfeits.  
More precisely, the set of telomeric extremities is mutable while, per definition, each derived genome shares the same set of non-telomeric extremities. 
To this end, our ILP makes use of additional variables that indicate the \emph{presence} of an extremity in a derived genome that is part of the solution (cf.~D.06).
Components corresponding to circular singletons are counted by binary variables specified in~D.07. 
Last, we also count the number of linear chromosomes in derived genomes of the solution (cf.~D.08).

\smallskip
Valid solutions to Problem~\ref{prb:wdeg_dcj} for each pair of degenerate genomes $\{A, B\} \in E(\phylo)$ are guaranteed by constraints~C.01-09.
The presence of each node associated with a non-telomeric extremity in capped relational diagram $R_\circ(A', B', \family')$ is ensured by setting the corresponding variable $o_v$ for each $v$ in $V^{AB}$ to one. 
Constraints~C.02 enforce that each connected component in $R_\circ(A', B', \family')$ represents an alternating cycle, where each adjacency edge is followed by either an extremity or indel edge.  
The next constraint, C.03, implements the definition of a family assignment which specifies that head and tail of marker in $\{A, B\}$-resolved assignment $\family'$ must belong to the same family. 
Here, we denote by \emph{sibling} a pair of edges $\{g^\et, h^\et\}, \{g^\eh, h^\eh\} \subseteq E_\adj^{AB}$, such that $\{g^\et, g^\eh\}, \{h^\et, h^\eh\} \in \markers$. 

Variables $y^{AB}$ label each cycle in relational diagram $R_\circ(A', B', \family')$ by a number (cf.~C.04). This number will be zero, if the cycle contains one or more indel edges (cf.~C.05). 
Otherwise, the number will correspond to the smallest index of any node in the cycle (cf.~C.06).
Constraints~C.07 label runs of genome $A$ as zero and runs of genome $B$ as one. 
Constraint~C.08 enforces that the indicator of a transition edge $t_{uv}^{AB}$ corresponding to edge $\{u, v\} \in E^{AB}$ is set to one if (i) the edge is part of relational diagram $R_\circ(A', B', \family')$ and (ii) its incident nodes have unequal run labels.

Constraint C.09~sets the indicator $s_C$ for a connected component $C$ to one if $C$ is a circular singleton. 
This entails the prerequisite that the set of all candidates of circular singletons $\mathcal C^{AB}$ must be known.
The construction of this set is explained further below. 

Constraints~C.10 and C.11~are \emph{optional} in the sense that they are not required to construct a valid solution to SPP-DCJ. 
However, these constraints reduce the search and/or solution space, which helps the ILP solver in reducing the computation time. 
Constraint C.10 limits the choice of transition edge between to neighboring indel runs to an adjacency edge of genome $A$. 
Last, constraints C.11 let the solver know that valid genomes contain an even number of telomeric extremities. 

\subsection{Identifying candidates of circular singletons}
For the construction of the ILP described in Algorithm~\ref{alg:ilp} the set of circular singleton candidates  $\mathcal C^{AB}$, $\{A, B\} \in E(\phylo)$ must be known. 
The number of candidates depends heavily on the surfeit of degenerate genomes $A$ and $B$ and is bounded by the number of ordered partitions of edge set $E_\id^A$ and $E_\id^B$, respectively. 
These numbers are also known as \emph{ordered Bell numbers}. 
In practice, the size of $\mathcal C^{AB}$ is small, because degenerate genomes have typically low surfeit. 
We construct the set of circular singleton candidates by traversing the multi-relational diagram, thereby identifying alternating cycles of adjacency and indel edges. 

\subsection{Reducing the search space of optimal path recombinations}
A substantial factor that impacts the running time are large marker families and the number telomeres in degenerate genomes~\cite{BohnenkamperBDS20}. 
Large numbers of telomeres act in similarly to large marker families, as in the capping strategy used in the construction of the multi-relational diagram $\MR(A, B, \family)$  extremity edges are drawn between all telomeric extremities of the degenerate genomes $A$ and $B$. 
This is to include path recombinations in the optimization that may further minimize the DCJ-indel distance. 
If path recombinations can be determined beforehand, then the search space of optimal path recombinations can be reduced by omitting non-optimal recombinations. 
However, paths used in the solution of SPP-DCJ are often unknown, as they depend on the choice of indel and extremity edges. 
Still, for any two telomeres, it is possible to identify the subgraph that spans all possible alternating paths that connect them. 
We then use these subgraphs to classify the telomere pairs into two groups: those, contained in indel-free subgraphs, and those, contained in indel-enclosing subgraphs.
As the former group does not contribute to the indel-penalty, the optimization of their recombinations will maximize the count of indel-free cycles. 
Conversely, the optimization of indel-enclosing cycles does not affect that count, but will only reduce the indel-penalty. 
So path recombinations within the two classes of telomeres can be optimized independently. 

We identify the class membership of telomeres using a depth-first search (DFS) strategy, which, for a given telomere $t \in V^{AB}$, simultaneously identifies all possible telomeres that can be recombination partners of $t$. 
While traversing the graph to find these partners, we also record the occurrence of indel edges. 
The procedure is repeated for all telomeres. 
Overall, the classification completes in $O(|V^{AB} \cap T| \cdot |E^{AB}|)$ time.
Identified indel-free recombination paths of telomeric extremities from opposite genomes can be directly connected by an extremity edge. 
However, indel-free paths between telomeres of the same genome and indel-enclosing path combinations of opposite genomes are connected by extremity edges in an all-vs-all manner. 
If the latter group has an unequal number of telomeres in $A$ and $B$, additional telomeres are added, as described above in the construction of the multi-relational diagram. 

\section{Results}\label{sec:results}

We implemented a Python program that constructs the ILP of Algorithm~\ref{alg:ilp} for a given input data set. The source and instructions for its usage is available at \url{https://github.com/danydoerr/spp_dcj}.
The ILP is solved with Gurobi~\cite{gurobi}. 

\subsection{Simulation experiments}

We evaluated our method on simulated data of moderate scale, using a phylogeny with 10 extant and 9 ancestral genomes. 
The genomes and their reconciled gene trees were generated with the genome evolution simulator ZOMBI~\cite{ZOMBI}. 
ZOMBI simulates intra-chromosomal evolution and supports a variety of evolutionary events, including inversions, transpositions, segmental duplications, and segmental deletions. 
To generate the data sets, we gradually increased the evolutionary scale (number of expected events per branch of the phylogenetic tree), thus creating series of data sets with increasing numbers of evolutionary events. 
We generated 80 data sets with varying numbers of above-mentioned events. 
Tables~\ref{tab:zombi_dataset1} and~\ref{tab:zombi_dataset2} in the Appendix display these numbers for two representative series of data sets.

Each simulated genome is associated with a node of the phylogeny and corresponds to a single circular chromosome comprised of \emph{ca.}~$1{,}000$ markers, the exact number depending on the rate of gene duplication and loss used in the simulation. 
We conducted three different experiments, each introducing a different type of noise in the data used as input to the ancestral genome reconstruction workflow.
In all these experiments, we used the true reconciled gene trees in the input data.
The rationale to rely on true reconciled gene trees is to avoid conflating the analysis of the SPP algorithm by factors that are external and therefore unrelated to the ILP when assessing its quality of reconstruction. 
We discuss this aspect further in conclusion.
At last, we set $\alpha=\frac{1}{2}$ and $\beta=\frac{1}{4}$ for calculating the weighted degenerate DCJ-indel distance in all experiments. 

\emph{Random Noise.} In the first experiment, we tested our method for susceptibility of biased and unbiased noise:
We generated ancestral degenerate genomes by augmenting the true ancestral genomes of our simulated data sets with random ancestral adjacencies of two kinds: uniform random adjacencies, sampled uniformly from the space of all possible adjacencies of a genome, and \emph{adversarial} random adjacencies formed between extremities of pairs of genes whose families already form adjacencies in the parental degenerate genome.  
Adversarial adjacencies are particularly detrimental to parsimony-based reconstruction as they are natural candidates to be conserved along a branch of the phylogeny. 
Altogether, we generated three batches, each comprising 480 data sets, with $0\%$, $50\%$, and $100\%$ of the added random adjacencies being adversarial. 
Within each batch, we modulated the surfeits of the degenerate ancestral genomes within the range of $1.2$ to $3$ by adding varying numbers of random adjacencies. 
In that experiment all adjacencies received weight $1$.

Except for $26$ data sets, SPP-DCJ perfectly recovered the true adjacencies in all three batches.
Of the data sets with incorrectly reconstructed adjacencies, $1$ occurred in a data set whose degenerate genomes had surfeit $2$, while the remaining had surfeit $3$.
Still, SPP-DCJ erred only on very few adjacencies, scoring after all 99.999\% and 99.7\% in mean precision and recall, respectively. 
This baseline experiment shows that a parsimony approach in the evolutionary model implemented in SPP-DCJ is able to recover the true evolutionary signal even in the presence of significant levels of noise, including adversarial noise explicitly designed at simulating parsimonious evolution of false positive adjacencies.

\emph{Principled Noise.} In our second set of experiments, we reconstructed ancestral genomes based again on including the true ancestral adjacencies in the ancestral degenerate genomes, but augmented by adjacencies recovered by DeCoSTAR~\cite{DucheminAP17}.
We used DeCoSTAR on the true reconciled gene trees and subsequently derived ancestral adjacencies from the returned adjacency evolution forests. 
This experiment thus includes a crucial step of the ancestral reconstruction workflow described in Figure~\ref{fig:workflow}, that is the inference of ancestral adjacencies based on local parsimony (at the level of pairs of gene families) that is more likely to generate false positive adjacencies (see~\cite{tannier:hal-02535466} for similar experiments with this framework), although keeping the true adjacencies in the input too.
In the calculation of the weighted degenerate DCJ-indel distance we weighted true adjacencies by one and DeCoSTAR adjacencies by its own assigned weights. 
In 78 out of 80 experiments, SPP-DCJ obtained a perfect score in recovering the true adjacencies. 
In the remaining two experiments, 2 (resp. 4) adjacencies were incorrectly selected by SPP-DCJ. 
This indicates that in rare cases, DeCoSTAR introduces bias that negatively impacts the accurate reconstruction of ancestral gene orders. 

\begin{figure}[tb]
    {\centering
    \includegraphics[width=0.5\columnwidth]{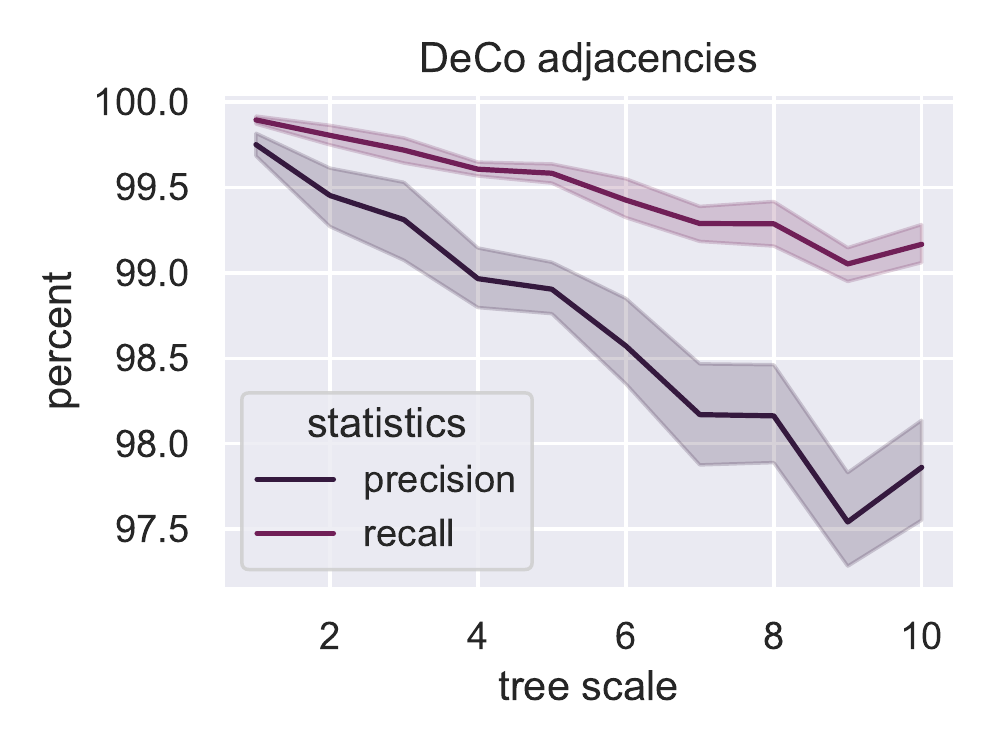}%
    \includegraphics[width=0.5\columnwidth]{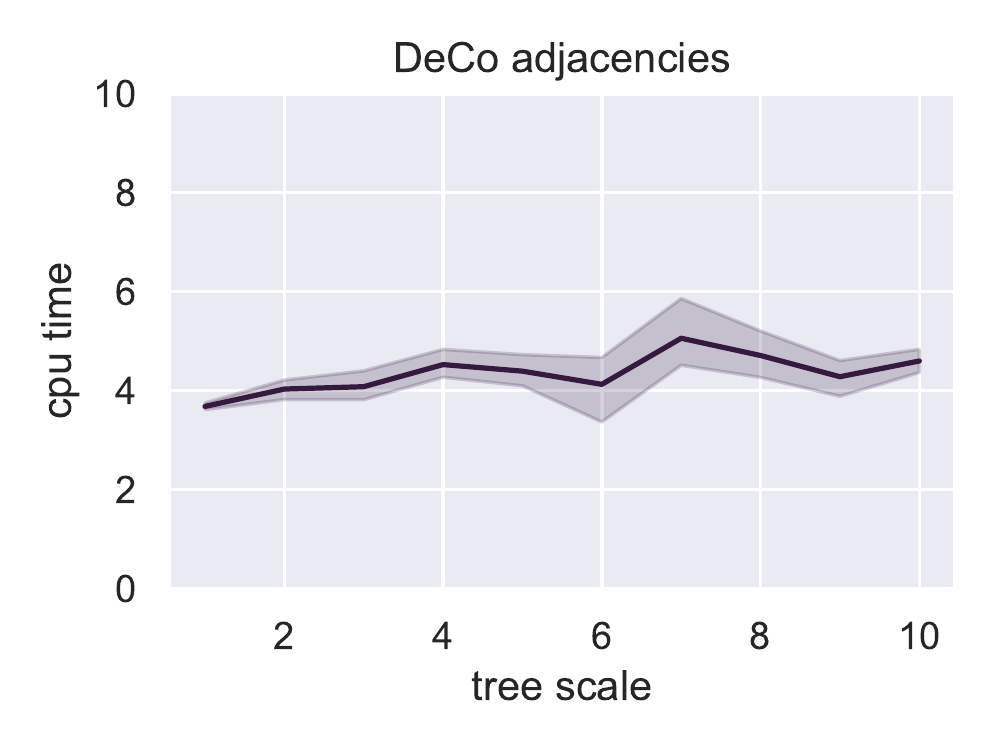}


    \includegraphics[width=0.5\columnwidth]{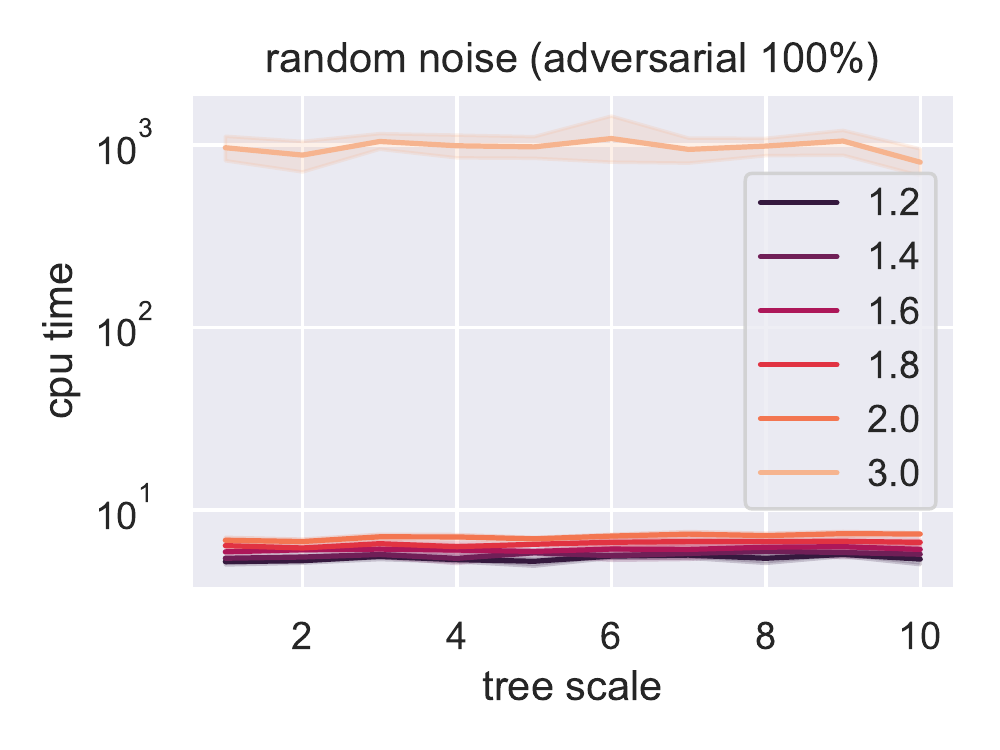}%
    \includegraphics[width=0.5\columnwidth]{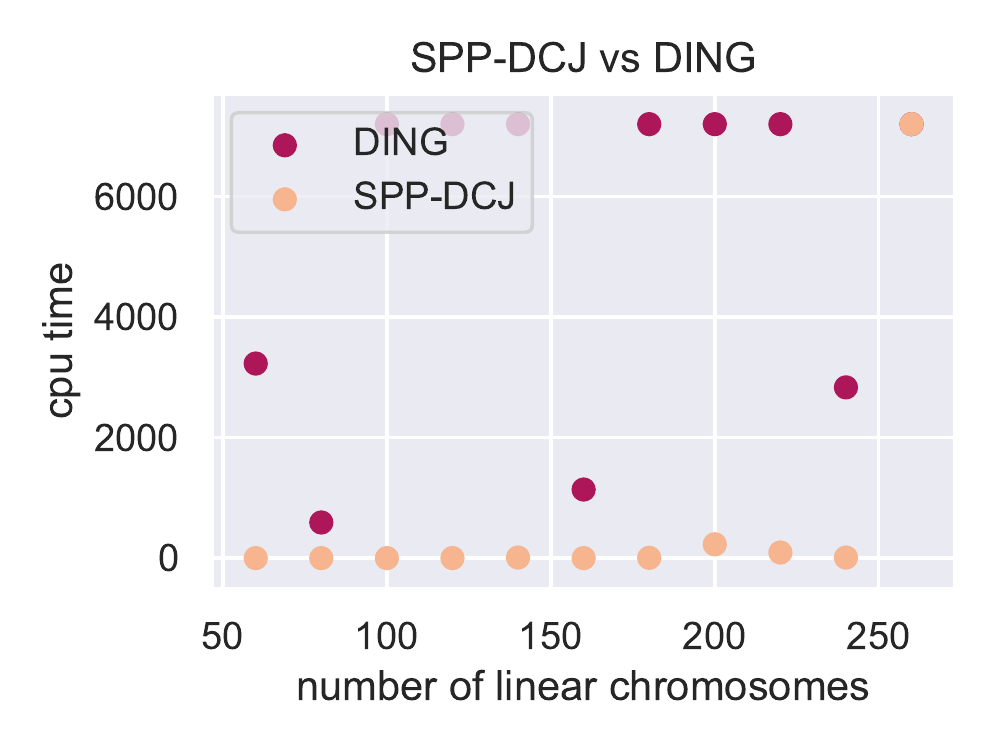}\\}
    
    \caption{
    \textbf{(Top left)} Precision and recall of reconstructed ancestral adjacencies in experiments with ancestral degenerate genomes defined by DeCoSTAR's inferred adjacencies.
    \textbf{(Top right)} Runtime of experiments using  DeCoSTAR's inferred adjacencies.
    \textbf{(Bottom left)} Runtime of SPP-DCJ on noisy data with varying surfeit and 100\% adversarial noise as a function of tree scale.
    \textbf{(Bottom right)} Runtime comparisons of SPP-DCJ and DING for pairwise distance computations.
    }\label{fig:results}
\end{figure}

\emph{DeCoSTAR Reconstruction.} In the last experiment we scrutinize the second part of the reconstruction workflow by performing the last three out of five steps (see Figure~\ref{fig:workflow}) as if the underlying data were not simulated, but represented  actual biological data. 
In other words, we applied SPP-DCJ only on adjacencies inferred by DecoSTAR. 
Overall, the precision and recall of adjacencies in the derived genomes of SPP-DCJ do not fall below 97\% and 98.7\%, respectively. 
The plot on the top left of Figure~\ref{fig:results} visualizes both statistics for all samples as a function of the scale of the phylogeny. 
As expected, the quality of reconstruction declines with increasing scale, \emph{i.e.}, numbers of evolutionary events.

\subsection{Runtime analysis.}
In all experiments described above except those cases of the first experiment where degenerate genomes were excessively large (surfeit $3$), Gurobi needed  only few seconds to find an optimal solution for SPP-DCJ (cf.~Figure~\ref{fig:results} top right and bottom left).
For the data sets with large degenerate genomes $3$, the program completed on average after 16 minutes, as indicated by bottom left plot of Figure~\ref{fig:results}. 
This is impressive as SPP-DCJ must calculate the DCJ-indel distance for 18 pairs of degenerate genomes simultaneously. 
Just as its predecessor, the runtime of SPP-DCJ is only weakly affected by genome size. 
To prove this point, we generated a series of data sets with genomes comprising $10{,}000$ markers and numbers of evolutionary events comparable to those described in Table~\ref{tab:zombi_dataset2} in the Appendix. 
SPP-DCJ found optimal solutions in each run after 47 CPU seconds.

SPP-DCJ extends DING~\cite{BohnenkamperBDS20} from pairwise distance calculations to the here introduced small parsimony setting. It outperforms its predecessor for genomes with many linear chromosomes. 
To illustrate this, we ran an experiment where we simulated the evolution of genomes with many linear chromosomes (between 60 and 260), close in nature to haplotype-resolved human genomes, or partially assembled genomes.
Our approach in handling optimal recombinations of indel-free paths tremendously reduces the running time.
In the experiment, we limited the running time of the pairwise distance solver described in~\cite{BohnenkamperBDS20} to two hours. 
While SPP-DCJ completes all but one pairwise calculations within a few seconds, DING reaches the runtime limit in six out of the eleven runs (cf.~bottom right plot of Figure~\ref{fig:results}).

\section{Conclusion}\label{sec:conclusion}
In this work we introduced a novel method for solving the SPP in the DCJ-indel model, within an ancestral reconstruction workflow based on the seminal work of Sankoff and El-Mabrouk~\cite{Sankoff2000} and taking advantage of recent advances in the field of phylogenomics~\cite{DucheminAP17}. 

Our experiments on simulated data of moderate size provides a proof of concept that SPP-DCJ can reconstruct efficiently accurate ancestral gene orders.
Assessing the performance of SPP-DCJ on data of larger scale is a natural avenue for further work, as is a more thorough exploration of the impact of errors in reconciled gene trees~\cite{tannier:hal-02535466}.
The joint reconstruction of ancestral gene orders and extant scaffolds using DeCoSTAR has recently been applied with good results to a large-scale datasets of mosquito genomes~\cite{ANOPHELESc}.
SPP-DCJ naturally lends itself to such an application as degenerate extant gene orders can be considered by the method; the application of SPP-DCJ to comparative scaffolding is thus a promising research direction. 
Last, the SPP-DCJ algorithm can work with non-treelike phylogenies. 
This makes our method applicable to population genomics studies, where introgression events introduce reticulation nodes in the species phylogeny. 


\bibliographystyle{acm}
\bibliography{main-arxiv}

\appendix

\newpage
\section{Simulation Details}
\begin{table}[h]
    \caption{\emph{Data set 1}: Total numbers of evolutionary events produced by ZOMBI run on tree of 10 extant species with parameter settings: \emph{genome size}: 1000, \emph{duplication}: $2$, \emph{dupl. extension}: $0.5$, \emph{loss}: $2$, \emph{loss extension}: $0.5$, \emph{inversion}: $2$, \emph{inversion extension}: $0.05$, \emph{transposition}: $2$, \emph{transposition extension}: $0.05$, \emph{origination}: $0$.}\label{tab:zombi_dataset1}

    {\centering
        \renewcommand{\arraystretch}{1.5}
    \scriptsize{
        \begin{tabular}{p{1cm}p{1cm}p{1cm}p{1cm}p{1cm}p{1cm}p{1.2cm}}
    \toprule
    \textbf{Tree scale}  &  \textbf{Dup. events}  &  \textbf{Dup. genes}  &  \textbf{Loss events}  &  \textbf{Lost genes}  &  \textbf{Inver-sions}  &  \textbf{Transpo-sitions} \\
    \midrule
    1 &           4 &         10 &            8 &          14 &           4 &               4 \\
    2 &           9 &         16 &           13 &          25 &          16 &               8 \\
    3 &           5 &         15 &           17 &          35 &          13 &              12 \\
    4 &          26 &         47 &           28 &          60 &          18 &              15 \\
    5 &          29 &         51 &           35 &          72 &          18 &              28 \\
    6 &          33 &         80 &           35 &          70 &          27 &              26 \\
    7 &          32 &         69 &           27 &          57 &          32 &              30 \\
    8 &          50 &         89 &           34 &          63 &          47 &              34 \\
    9 &          40 &         90 &           48 &         102 &          48 &              45 \\
    10&          55 &        107 &           52 &         117 &          55 &              44 \\
    \bottomrule
    \end{tabular}}\\}
\end{table}

\begin{table}[h!]
    \caption{\emph{Data set 2}: Total numbers of evolutionary events produced by ZOMBI run on tree of 10 extant species with parameter settings: \emph{genome size}: 1000, \emph{duplication}: $2$, \emph{dupl. extension}: $0.8$, \emph{loss}: $2$, \emph{loss extension}: $0.8$, \emph{inversion}: $2$, \emph{inversion extension}: $0.5$, \emph{transposition}: $2$, \emph{transposition extension}: $0.5$, \emph{origination}: $0$.}\label{tab:zombi_dataset2}

    {\centering
        \renewcommand{\arraystretch}{1.5}
    \scriptsize{
        \begin{tabular}{p{1cm}p{1cm}p{1cm}p{1cm}p{1cm}p{1cm}p{1.2cm}}
    \toprule
    \textbf{Tree scale}  &  \textbf{Dup. events}  &  \textbf{Dup. genes}  &  \textbf{Loss events}  &  \textbf{Lost genes}  &  \textbf{Inver-sions}  &  \textbf{Transpo-sitions} \\
    \midrule
    1  &           3 &          5 &            4 &           4 &           6 &               3 \\
    2  &           4 &          5 &            9 &          12 &           7 &              10 \\
    3  &          15 &         18 &           17 &          22 &          11 &              22 \\
    4  &          21 &         29 &           17 &          24 &          17 &              21 \\
    5  &          24 &         33 &           30 &          37 &          22 &              18 \\
    6  &          35 &         48 &           39 &          44 &          23 &              37 \\
    7  &          52 &         67 &           26 &          30 &          40 &              42 \\
    8  &          39 &         46 &           40 &          51 &          37 &              39 \\
    9  &          45 &         60 &           45 &          59 &          53 &              38 \\
    10 &          43 &         57 &           47 &          63 &          53 &              54 \\
    \bottomrule
    \end{tabular}}\\}
\end{table}

\end{document}